\documentclass[12pt]{article}
\usepackage{graphicx,epsfig,epsf}

\usepackage{graphicx}
\usepackage{scicite}
\usepackage{times}

\def\ltsima{$\; \buildrel < \over \sim \;$}
\def\simlt{\lower.5ex\hbox{\ltsima}} 
\def\gtsima{$\; \buildrel > \over \sim \;$}
\def\simgt{\lower.5ex\hbox{\gtsima}} 

\def\deg{\hbox{$^\circ$}}
\def\etal{{\it et al.}}
\def\Ms{$M_\odot$}

\def\p0{$\pi^{\rm 0}$}
\def\Angst{$\buildrel _{\circ} \over {\mathrm{A}}$}
\def\Fermi{\textit{Fermi}}
\def\Swift{\textit{Swift}}

\def\aj{{\it Astron.~J.}}
\def\aph{{\it Astropart.~Phys.}}
\def\apj{{\it Astrophys.~J.}}
\def\apjl{{\it Astrophys.~J.~Lett.}}
\def\apjs{{\it Astrophys.~J.~Supp.}}

\def\ar{{\it Astron.~Rep.}}
\def\atel{{\it The Astronomer's Telegram}}
\def\mnras{{\it Mon.~Not.~R.~Astron.~Soc.}}
\def\nat{{\it Nature}}
\def\nimpa{{\it Nucl.~Instrum.~Methods~Phys.~Res.,~Sect.~A}}
\def\pasj{{\it Publ.~Astron.~Soc.~Japan}}
\def\pasp{{\it Publ.~Astron.~Soc.~Pac.}}
\def\rmp{{\it Rev.~Mod.~Phys.}}

\newenvironment{sciabstract}{ \begin{quote} \bf} {\end{quote}}

\newcounter{lastnote}

\title{Gamma-ray Emission Concurrent with the Nova in the Symbiotic 
Binary V407 Cygni}

\author{The Fermi-LAT Collaboration\footnote{All authors with their 
affiliations appear at the end of this paper.} 
}

\date{submitted 19 May 2010; accepted 12 July 2010}

\begin{document}

\baselineskip24pt

\maketitle 

\begin{sciabstract} 

Novae are thermonuclear explosions on a white dwarf surface fueled by 
mass accreted from a companion star.  Current physical models posit that 
shocked expanding gas from the nova shell can produce X-ray emission but 
emission at higher energies has not been widely expected. Here, we 
report the \Fermi\ Large Area Telescope detection of variable 
$\gamma$-ray (0.1--10 GeV) emission from the recently-detected optical 
nova of the symbiotic star V407 Cygni. We propose that the material of 
the nova shell interacts with the dense ambient medium of the red giant 
primary, and that particles can be accelerated effectively to produce 
\p0\ decay $\gamma$-rays from proton-proton interactions. Emission 
involving inverse Compton scattering of the red giant radiation is also 
considered and is not ruled out. \end{sciabstract}

V407 Cygni (V407 Cyg) is a binary system consisting of a Mira-type 
pulsating red giant (RG) with a white dwarf (WD) companion; these 
properties place it among the class of symbiotic binaries \cite{ken92}.  
Historically, although one of the more active symbiotic systems, V407 
Cyg showed an optical spectrum in quiescence dominated by the Mira-like 
RG (M6 III) and only weak emission lines, e.g., \cite{mun90}. Its 
infrared continuum (consistent with a dusty wind) and maser emission 
\cite{deg05} are detected at levels similar to other symbiotic Miras 
[e.g., R Aqr \cite{ivi94}]. One outstanding anomaly of V407 Cyg is a 
strong Li I $\lambda$6707 line indicative of an overabundance of Li 
relative to normal Mira red giants \cite{tat03,shu07}. Based on the 
745-day pulsation period of the RG \cite{mei66} and the Mira 
period-luminosity relation \cite{gla82}, we adopt the distance, $D=2.7$ 
kpc, estimated as the mean derived from photometry in three 
near-infrared bands assuming an extinction, $E_{\rm B-V}=0.57$ 
\cite{mun90}.

A nova outburst from V407 Cyg was detected on 10 March 2010 
\cite{nis10}; it had a magnitude approximately 6.9 in an unfiltered CCD 
image obtained at 19:08 UT. Subsequent densely sampled observations show 
the outburst was followed by a smooth decay, though the precise epoch of 
the nova is formally uncertain by up to 3 days due to the time gap from 
the pre-outburst image (Fig.~1). Monitoring of the source over the past 
two years indicates pre-outburst magnitude values in the range 9--12 
[see the supporting online material (SOM)]. V407 Cyg has been monitored 
optically for decades and has shown earlier signs of optical brightening 
on month-timescales by 1--2 magnitudes in the $B$ and $V$ bands (around 
1936 and 1998) from typical $V$ band magnitudes of 13--16 
\cite{hof49,kol03,mun90}, but the magnitude of the recent nova was 
unprecedented.

Here we report on a high-energy $\gamma$-ray source (Fig.~2) 
positionally coincident with V407 Cyg detected after the nova 
\cite{che10} during routine automated processing of all-sky monitoring 
data from the \Fermi\ Large Area Telescope (LAT) \cite{atw09}. A 
$\gamma$-ray light curve (one-day time bins) of this source generated 
from an analysis of all LAT data reveals that the first significant 
detection (4.3$\sigma$) was in fact on 10 March indicating the 
$\gamma$-ray activity began on the same day as the reported optical 
maximum of V407 Cyg (Fig.~1, SOM). The observed 10 March flux is up to a 
factor of 3 larger than the one-day upper limits (unless otherwise 
noted, 95$\%$ confidence limits are reported throughout) on the 
pre-outburst days. To further isolate the onset of detectable 
$\gamma$-ray emission, we divided the 10 March data into 6-hour 
intervals and the first indication of a signal was a marginal detection 
in the 12h--18h bin (2.8$\sigma$). This was followed by a highly 
significant detection (at 5.7$\sigma$) in the last 6-hr bin (18h--24h) 
which had a peak flux that was a factor of $\simgt$3 larger than that of 
the marginal detection and the limits from earlier in the day (SOM). The 
initial detection of the $\gamma$-ray source by the \Fermi-LAT in the 
latter 1/2-day of 10 March is consistent with the time of the optical 
nova discovery.

The peak flux in $\gamma$-rays (defined on one-day segments) was 
observed between 13--14 March, 3--4 days after the initial $\gamma$-ray 
detection, and with a factor of 2 greater flux ($9 \times 10^{-7}$ 
photons cm$^{-2}$ s$^{-1}$). Analyzing data up to mid-April, the last 
day with a significant detection ($>$3$\sigma$) of the variable 
$\gamma$-ray source is 25 March, amounting to a total lifetime of 
activity of two weeks. Defining an `active' period\cite{onset} from 10 
March 18:00 to 29 March 00:00, we obtained a $\gamma$-ray position 
(Fig.~2) from the cumulative exposure that is 0.040\deg\ offset from the 
optical position of V407 Cyg, which is within the LAT 95$\%$ confidence 
circle (radius = 0.062\deg). The average spectral energy distribution 
(SED) of the \Fermi-LAT $\gamma$-ray source during the defined active 
period can be described with an exponentially cut off power-law model 
(SOM) with a flux ($>$100 MeV) of $(4.4 \pm 0.4) \times 10^{-7}$ photons 
cm$^{-2}$ s$^{-1}$ (overall source significance of 18.1$\sigma$). A 
likelihood ratio test demonstrates that the addition of the exponential 
cut off improves the fit at the 4.9$\sigma$ level compared to a simpler 
single power-law model. We find no evidence for spectral variability 
over the duration of the active $\gamma$-ray period (SOM). Analyzing 
data from the two weeks (29 March to 12 April) following the active 
period collectively gives a significance of only 1.6$\sigma$ (flux upper 
limit of $0.8 \times 10^{-7}$ photons cm$^{-2}$ s$^{-1}$), indicating 
that the flux has declined below detectability. Overall, the 
$\gamma$-ray source is brightest at earlier times, consistent with the 
optical behavior of V407 Cyg. The coincident localization and the 
observed correlated variability imply that the optical nova is the 
source of the variable $\gamma$-ray flux.

Nuclear $\gamma$-ray lines and continuum emission from novae at $\simlt 
1$~MeV energies have been considered \cite{her08}, but the \Fermi-LAT 
detection of V407 Cyg shows unequivocally that novae can generate 
high-energy ($>$100 MeV) $\gamma$-rays. The $>$100 MeV $\gamma$-ray 
luminosity, its spectrum, and light curve can be understood broadly as 
consequences of shock acceleration taking place in a nova shell.  Such a 
nova shell is produced by thermonuclear energy release on the WD and 
initially expands freely into a very dense medium consisting of the RG 
wind and atmosphere present in the binary system prior to the nova. The 
radio source detected from V407 Cyg over 22--28 March 
\cite{nes10,bow10}, and subsequent imaging which revealed the emission 
to be predominantly extended at few milli-arcsecond resolution 
\cite{gir10}, is consistent with a picture of an extended shell as was 
found in the 2006 RS Oph nova outburst \cite{obr06,rup08}.

An initially spherical shell can sweep up mass from the companion RG 
wind asymmetrically and will reach a deceleration phase during which it 
expands adiabatically \cite{sed59} with different temporal behavior in 
different directions from the WD center. Fermi acceleration of protons 
and electrons takes place in the outgoing nova shock during both the 
free expansion phase and deceleration phase, and we show that the 
measured $\gamma$-ray spectrum can be explained by \p0\ decay 
$\gamma$-rays from proton-proton ($pp$) collisions or inverse Compton 
(IC) scattering of infrared photons from the RG by electrons. In both 
these scenarios, the $\gamma$-ray light curve in conjunction with the 
delayed rise of the X-ray flux can be explained qualitatively as a 
geometrical effect of the nova shell evolution.

The measured optical peak magnitude $\approx 7$ of V407 Cyg \cite{nis10} 
over one day, implies an energy release of $\ge 3\times 10^{42}$~ergs at 
visible frequencies (after extinction correction).  The kinetic energy 
of the ejecta in the nova shell, ${\cal E}_{\rm k} \sim 10^{44}$~ergs, can be 
estimated assuming a nova shell mass, $M_{\rm ej}\sim 10^{-6} \, 
M_\odot$ [which is a plausible value for a massive WD, $>$1.25\Ms\ 
\cite{sta00,yar05}], and the velocity, $v_{\rm ej} = 3200 \pm 
345$~km~s$^{-1}$, inferred from an H$\alpha$ line width measurement on 
14 March (SOM). The velocity of the outgoing shock wave is initially 
$v_{\rm sh}\sim v_{\rm ej}$. The onset of the deceleration phase and 
subsequent evolution of the shock wave are determined by the density of 
the RG stellar wind and atmosphere surrounding the nova shell, which 
depends on two poorly constrained parameters -- the WD-RG separation 
($a$) and the RG mass loss rate in the wind (${\dot M}_{\rm w}$).  As we 
will discuss shortly, the detection of the $\gamma$-ray flux within one 
day of the optical detection of the nova, a peak duration of 3-4 days, 
and subsequent decay within 15 days after the optical nova can be 
modeled as emission from the nova shell in a dense environment and 
mostly from along the WD-RG axis assuming an inverse-square law wind 
density profile from the RG center, with $a\sim 10^{14}$~cm, and ${\dot 
M}_{\rm w}\sim 3\times 10^{-7} \, M_\odot$~yr$^{-1}$. Both these values 
are a factor $\approx 2$ smaller than those suggested previously for the 
V407 Cyg system \cite{mun90}.

The density of particles in the RG wind is $n(R) = {\dot M}_{\rm w} \, 
[4\pi \, (R^2 + a^2 -2 a R \cos\theta) \, v_{\rm w} \, {\bar m}]^{-1}$.  
Here, the RG wind velocity, $v_{\rm w} \approx 10$~km s$^{-1}$, is based 
on optical spectra (SOM), the mean particle mass is ${\bar m} \approx 
10^{-24}$~g, $R$ is the distance from the WD center, and $\theta$ is the 
polar angle relative to the WD center.  The energy density in the RG 
radiation field is similarly, $u_{\rm IR}(R) = L_{\rm IR} \, [4\pi (R^2 
+ a^2 -2 a R \cos\theta) \, c]^{-1}$. The RG luminosity, $L_{\rm IR} 
\approx 10^4L_\odot$ \cite{mun90,kol03}, is consistent with a spectral 
fit to post-nova infrared measurements with a temperature of $\approx 
2500$~K (SOM).  Near the WD surface ($R\approx 0.01~R_\odot$), these 
densities are $n \sim 10^8$~cm$^{-3}$ and $u_{\rm IR} \sim 
0.01$~erg~cm$^{-3}$, and increase by up to an order of magnitude when 
the nova shell approaches the RG surface (i.e., along $\theta\approx$ 
0\deg) at a radius $r_{\rm RG} \approx 500~R_\odot$.  An equipartition 
of the energy density in the magnetic field expected to arise from 
turbulent motions in the wind, to the thermal energy density in the RG 
wind with temperature $T_{\rm w} \approx 700$~K \cite{mun90} gives a 
mean magnetic field, $B_{\rm sh}(R) = [32\pi \, n(R) \, k \, T_{\rm 
w}]^{1/2} \sim 0.03$~G in the shock wave when it is near the WD.  
Electrons and protons can be accelerated efficiently in this magnetic 
field \cite{tat07} and interact with the surrounding RG wind particles 
and radiation.

The time scale for $pp$ interactions for a \p0\ model to produce 
$\gamma$-rays in the shock wave is $t_{\rm pp}\approx 1/[4 \, n(R) \, c 
\, \sigma_{\rm pp}] \sim 2.8 \times 10^{6}$~s when the nova shell is 
near the WD. Here, $\sigma_{\rm pp} \approx 3 \times 10^{-26}$~cm$^{2}$ 
is the $pp$ cross-section.  Thus, $t/t_{\rm pp} \sim 3\%$ of the protons 
can interact to produce \p0\ emission on a time scale, $t=1$ day.  In an 
IC scenario, the cooling time scale for electrons with energy $E_{\rm e} 
\approx 5$~GeV that up-scatter 2500~K photons to $\approx 100$~MeV is 
$t_{\rm IC} \approx (3/4) \, m_{\rm e}^2 \, c^3 \, [\sigma_{\rm T} \, 
E_{\rm e} \, u_{\rm IR}(R)]^{-1} \sim 3.1 \times 10^5$~s.  Thus, 
$t/t_{\rm IC}\sim 28\%$ of the electrons produce $\gamma$-rays 
efficiently in a time scale, $t=1$ day.  The efficiency for $\gamma$-ray 
production in both the \p0\ and IC models increases substantially in the 
part of the nova shell that expands towards the RG ($\theta \sim$ 0\deg) 
and reaches the deceleration phase, by accumulating RG wind and 
atmospheric material of mass equal to $M_{\rm ej}$, at a distance $\sim 
7 \times 10^{13}$~cm in about 2.5 days.  The efficiency decreases 
rapidly in the part of the shell that expands away from the RG ($\theta 
\ge$ 90\deg) because of a decreasing density in both the RG wind and 
radiation.

In our scenario, most of the $\gamma$-rays come from the part of the 
nova shell approaching the RG. This can qualitatively explain the basic 
features of the $\gamma$-ray light curve (Fig.~1): (i) its onset within 
days of the optical nova and the peak flux reached in three days due to 
an increasing efficiency for $pp$ interactions and an increasing volume 
of the shock-accelerated particles, and (ii) the decline in the flux 
after about five days due to weakening of the shock wave after reaching 
the deceleration phase.  A highly significant one-day detection 
($6.5\sigma$) of an increase in $\gamma$-ray flux 9 days after the nova 
discovery (Fig.~1) could be due to the nova shell hitting a part of the 
RG surface or a nearby remnant of high density with a size scale of 
$\sim 10^{13} \, (v_{\rm sh}/1500~{\rm km~s^{-1}}) \, (t/{\rm 
1~day})$~cm.

A representative \p0\ decay model is shown in Fig.~3 fitting the LAT 
data with a cosmic-ray proton spectrum in the form of an exponentially 
cut off power-law, $N_{\rm p} = N_{\rm p,0} \, (W_{\rm p} + m_{\rm p} \, 
c^2)^{-s_{\rm p}} \, e^{-W_{\rm p}/E_{\rm cp}}$, where $E_{\rm cp}$, 
$W_{\rm p}$, and $m_{\rm p}$ are the cutoff energy, kinetic energy, and 
mass of the proton, respectively.  The $\gamma$-ray spectrum is well 
reproduced with a spectral index, $s_{\rm p} = 2.15^{+0.45}_{-0.28}$, 
similar to the expected spectrum from the Fermi acceleration mechanism, 
with $E_{\rm cp} = 32^{+85}_{-8}$ GeV (1$\sigma$ uncertainties) (SOM).  
The total energy in $\gamma$-rays above 100~MeV integrated over the 
active period is ${\cal E}_\gamma\approx 3.6 \times 10^{41}$~erg. The 
total energy in protons is ${\cal E}_{\rm p} \approx \epsilon_{\rm 
p}^{-1} \langle E_{\rm p}/E_\gamma \rangle {\cal E}_\gamma$, where 
$\epsilon_{\rm p}$ is the mean efficiency for $pp$ interactions and 
$1/\langle E_{\rm p}/E_\gamma \rangle \approx 0.2$ is the mean fraction 
of proton energy transfer to $\gamma$-rays per interaction in the 
$E_\gamma \ge 100$~MeV range.  The ratio of the total energy in protons 
that produce $\gamma$-rays to the kinetic energy, ${\cal E}_{\rm 
p}/{\cal E}_{\rm k} \sim 9\%$, with $\epsilon_{\rm p} \sim 0.2$ averaged over 
the $\gamma$-ray source lifetime of 15 days and the whole nova shell, 
similar to 1--10\% estimated in supernova remnants \cite{casa}. The 
ratio, however, is larger when considering dominant $\gamma$-ray 
emission coming mostly from the part of the shell that expands towards 
the RG.

The leptonic model is represented in Fig.~3 as the total of the IC 
spectrum plus a small contribution from bremsstrahlung emission, that 
arises from scattering of electrons with protons of density $n(R)$ in 
the shock wave (SOM).  The exponentially cut off power-law electron 
spectrum, $N_{\rm e} = N_{\rm e,0} \, W_{\rm e}^{-s_{\rm e}} \, 
e^{-W_{\rm e}/E_{\rm ce}}$, where $E_{\rm ce}$ and $W_{\rm e}$ are the 
cutoff energy and kinetic energy of electrons (GeV), respectively, 
reproduces the $\gamma$-ray spectrum with $s_{\rm e} = 
-1.75^{+2.40}_{-0.59}$ and $E_{\rm ce}= 3.2^{+2.6}_{-0.1}$~GeV 
(1$\sigma$ uncertainties) (SOM).  The total number of electrons required 
in steady state is $N_{\rm e,0} \approx 4 \times 10^{42}$, with a mean 
energy of 8.7~GeV which is larger than $E_{\rm ce}$ because of the steep 
spectrum.  The total energy in electrons over 15 days of $\gamma$-ray 
emission is ${\cal E}_{\rm e} \approx 4 \times 10^{41}$~ergs, averaged 
over the nova shell.  Thus the total energy in electrons that produce 
$\gamma$-rays in the IC model is a small fraction of the kinetic energy 
in the shell (${\cal E}_{\rm e}/{\cal E}_{\rm k} \sim 0.4\%$).

X-ray emission detected from V407 Cyg with the \Swift\ X-ray Telescope 
(XRT) as early as three days after the onset of the optical nova 
(Fig.~1) is likely due to shock-heating of ambient gas 
\cite{sok06,dra09}. The X-ray flux starts rising significantly about two 
weeks after the nova, coinciding with when the $\gamma$-ray flux 
declines below detectability.  In our geometric scenario, the sharply 
rising X-ray flux is due to the increasing volume of shocked gas in the 
nova shell expanding in the direction away from the RG. The X-ray flux 
peaks about 30 days after the explosion, and its subsequent slow decline 
is consistent with the longer time scale of the deceleration phase.

The \Fermi-LAT detection of V407 Cyg was a surprise, and adds novae as a 
source class to the high-energy $\gamma$-ray sky.  The particle 
acceleration mechanism and the $\gamma$-ray emission scenarios outlined 
here require the mass donor to be a red giant, i.e., a nova in a 
symbiotic system. Interestingly, several symbiotic stars are known to be 
recurrent novae (i.e., systems observed to have undergone multiple 
thermonuclear runaways within the last century), and recurrent novae are 
often considered candidate progenitors of Type Ia supernovae 
\cite{woo06}.  V407 Cyg may also belong to this class of binaries, and 
we have adopted parameters that are consistent with such a 
classification in modeling the $\gamma$-ray emission.  These sources can 
in general have dramatic influence on the local interstellar medium and 
Galactic cosmic rays but few binary systems with a WD are known to have 
a similar environment, hence we expect $\gamma$-ray novae to be rare.



\begin{figure}[htbp]
  \begin{center}
    \includegraphics[width=16cm]{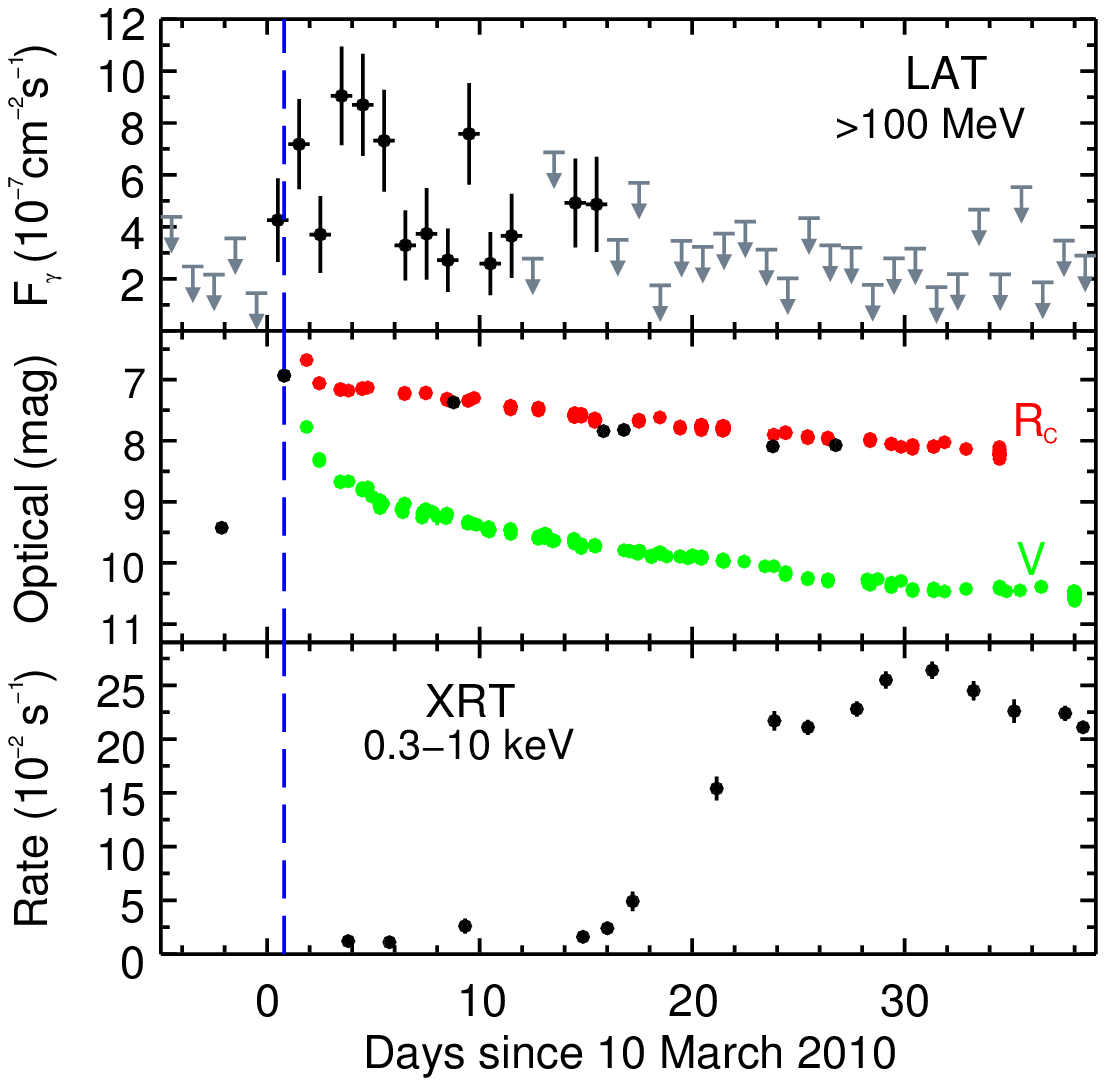}
  \end{center} {\bf Fig.~1.} Light curves of V407 Cyg in $\gamma$-rays 
from the \Fermi-LAT (top), optical (middle), and X-rays from \Swift\ 
(bottom). Vertical bars indicate 1$\sigma$ statistics errors for all 
data (the errors are smaller than the points in the optical). For the 
$\gamma$-ray data, gray arrows indicate 2$\sigma$ upper limits, and 
horizontal bars indicate the one-day binning. In the optical, unfiltered 
(black), $R_{\rm C}$ (red), and $V$ (green) band magnitudes are shown 
(SOM). The vertical dashed blue line indicates the epoch of the optical 
nova detection; the $\gamma$-ray peak occurred 3--4 days later. 
\end{figure}

\begin{figure}[htbp]
  \begin{center}
    \includegraphics[width=16cm]{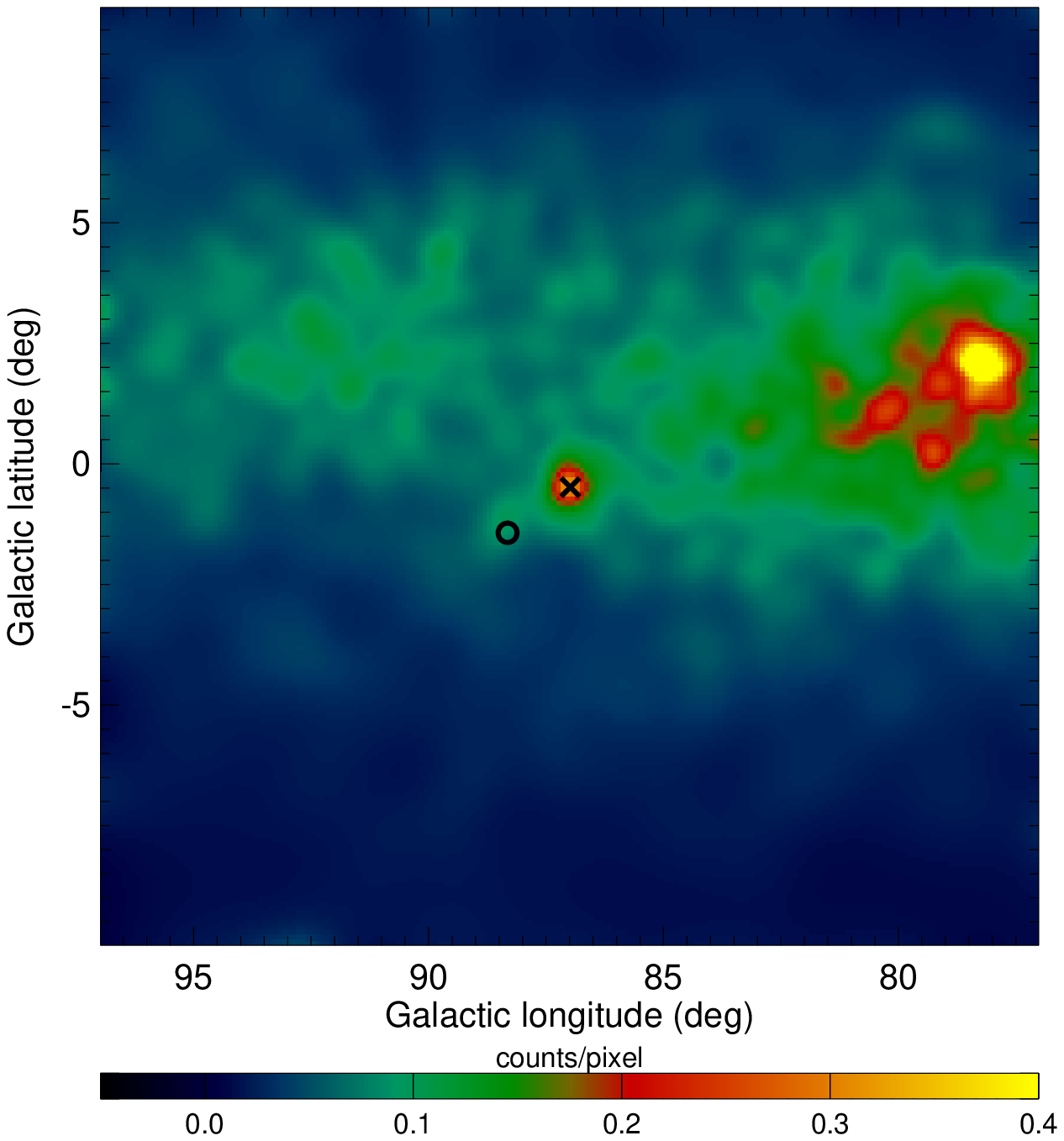}
  \end{center} {\bf Fig.~2.} \Fermi-LAT $\gamma$-ray counts (200 
MeV--100 GeV) map from 10 March 18:00 -- 29 March 00:00 2010 in the 
region around the position of the $\gamma$-ray nova source V407 Cyg 
(marked by black cross) at $l = 86.958\deg$, $b= -0.513\deg$ (R.~A.~= 
315.551\deg, Dec.~= +45.737\deg, J2000.0). The map was adaptively 
smoothed by imposing a minimum signal-to-noise ratio of 7. The closest 
known $\gamma$-ray source is contained in the 1st year LAT catalog 
(1FGL~J2111.3+4607; marked by the black circle) \cite{1fgl}, 
$\sim$1.5\deg\ away from the star's optical position. The bright source 
at ($l = 78.2\deg$, $b= +2.1\deg$) is LAT PSR~J2021+4026.\end{figure}

\begin{figure}[htbp]
  \begin{center}
    \includegraphics[width=15cm]{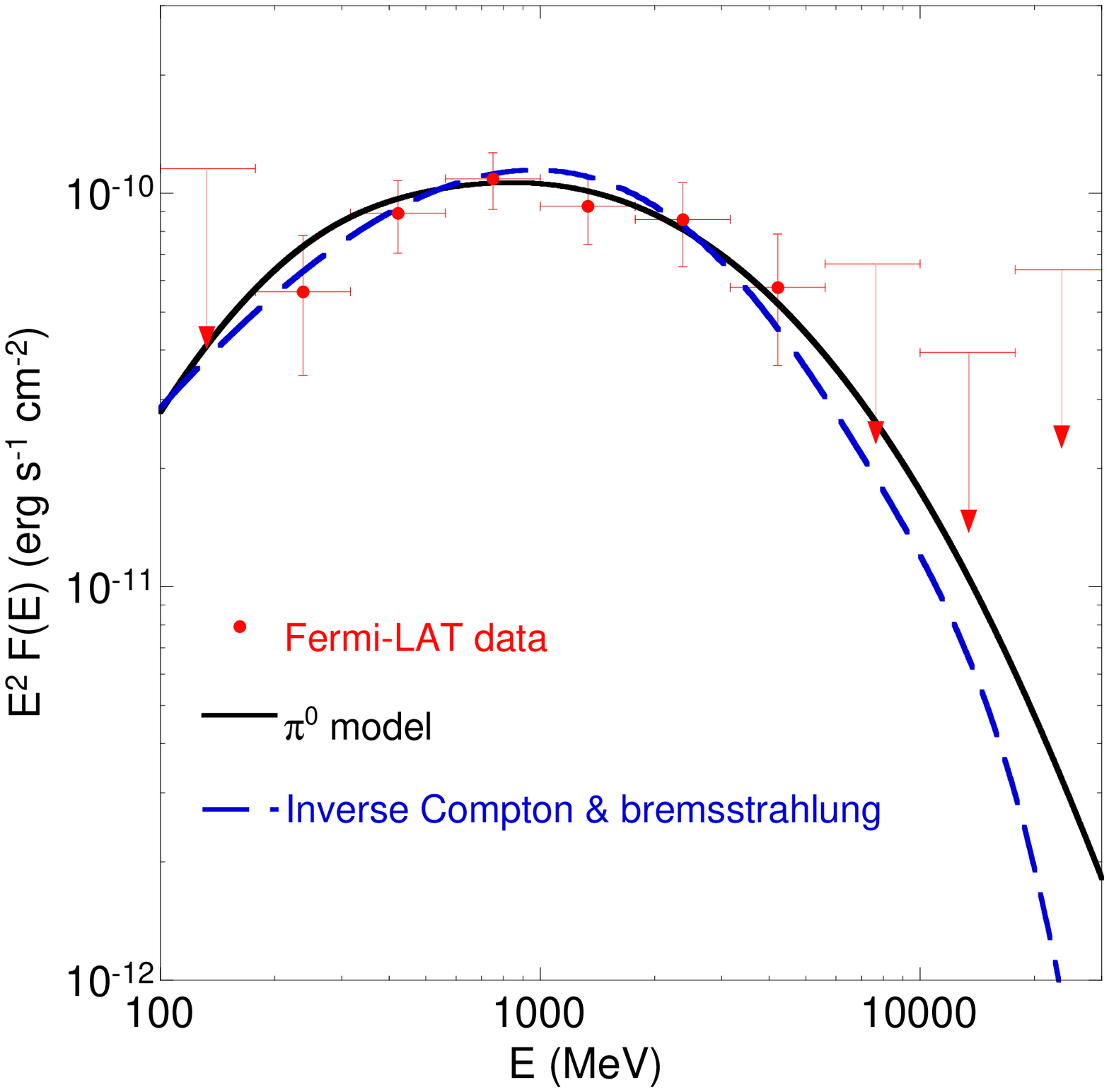}
  \end{center} {\bf Fig.~3.} SED of V407 Cyg in MeV/GeV $\gamma$-rays 
measured by the \Fermi-LAT over the period 10 March 18:00 -- 29 March 
00:00 2010. Vertical bars indicate 1$\sigma$ statistical errors, arrows 
indicate 2$\sigma$ upper limits, and horizontal bars indicate energy 
ranges. The best-fit \p0\ (black solid line) and leptonic (blue dashed 
line) models are indicated. \end{figure}

\clearpage

\noindent {\bf The Fermi-LAT Collaboration}\\
\noindent
A.~A.~Abdo$^{1,2}$, 
M.~Ackermann$^{3}$, 
M.~Ajello$^{3}$, 
W.~B.~Atwood$^{4}$, 
L.~Baldini$^{5}$, 
J.~Ballet$^{6}$, 
G.~Barbiellini$^{7,8}$, 
D.~Bastieri$^{9,10}$, 
K.~Bechtol$^{3}$, 
R.~Bellazzini$^{5}$
% ...
\begin{enumerate}
\item[1.] Space Science Division, Naval Research Laboratory, Washington, DC 20375, USA
\item[2.] National Research Council Research Associate, National Academy of Sciences, Washington, DC 20001, USA
\item[3.] W. W. Hansen Experimental Physics Laboratory, Kavli Institute for Particle Astrophysics and Cosmology, Department of Physics and SLAC National Accelerator Laboratory, Stanford University, Stanford, CA 94305, USA
\item[4.] Santa Cruz Institute for Particle Physics, Department of Physics and Department of Astronomy and Astrophysics, University of California at Santa Cruz, Santa Cruz, CA 95064, USA
\item[5.] Istituto Nazionale di Fisica Nucleare, Sezione di Pisa, I-56127 Pisa, Italy
\item[6.] Laboratoire AIM, CEA-IRFU/CNRS/Universit\'e Paris Diderot, Service d'Astrophysique, CEA Saclay, 91191 Gif sur Yvette, France
\item[7.] Istituto Nazionale di Fisica Nucleare, Sezione di Trieste, I-34127 Trieste, Italy
\item[8.] Dipartimento di Fisica, Universit\`a di Trieste, I-34127 Trieste, Italy
\item[9.] Istituto Nazionale di Fisica Nucleare, Sezione di Padova, I-35131 Padova, Italy
\item[10.] Dipartimento di Fisica ``G. Galilei", Universit\`a di Padova, I-35131 Padova, Italy
\item[11.] Istituto Nazionale di Fisica Nucleare, Sezione di Perugia, I-06123 Perugia, Italy
\item[12.] Dipartimento di Fisica, Universit\`a degli Studi di Perugia, I-06123 Perugia, Italy
\item[13.] Centre d'\'Etude Spatiale des Rayonnements, CNRS/UPS, BP 44346, F-30128 Toulouse Cedex 4, France
\item[14.] Department of Physics, Center for Cosmology and Astro-Particle Physics, The Ohio State University, Columbus, OH 43210, USA
\item[15.] Dipartimento di Fisica ``M. Merlin" dell'Universit\`a e del Politecnico di Bari, I-70126 Bari, Italy
\item[16.] Istituto Nazionale di Fisica Nucleare, Sezione di Bari, 70126 Bari, Italy
\item[17.] Laboratoire Leprince-Ringuet, \'Ecole polytechnique, CNRS/IN2P3, Palaiseau, France
\item[18.] Department of Physics, University of Washington, Seattle, WA 98195-1560, USA
\item[19.] Institut de Ciencies de l'Espai (IEEC-CSIC), Campus UAB, 08193 Barcelona, Spain
\item[20.] INAF-Istituto di Astrofisica Spaziale e Fisica Cosmica, I-20133 Milano, Italy
\item[21.] NASA Goddard Space Flight Center, Greenbelt, MD 20771, USA
\item[22.] Center for Research and Exploration in Space Science and Technology (CRESST) and NASA Goddard Space Flight Center, Greenbelt, MD 20771, USA
\item[23.] Department of Physics and Center for Space Sciences and Technology, University of Maryland Baltimore County, Baltimore, MD 21250, USA
\item[24.] George Mason University, Fairfax, VA 22030, USA
\item[25.] Laboratoire de Physique Th\'eorique et Astroparticules, Universit\'e Montpellier 2, CNRS/IN2P3, Montpellier, France
\item[26.] Department of Physics, Stockholm University, AlbaNova, SE-106 91 Stockholm, Sweden
\item[27.] The Oskar Klein Centre for Cosmoparticle Physics, AlbaNova, SE-106 91 Stockholm, Sweden
\item[28.] Royal Swedish Academy of Sciences Research Fellow, funded by a grant from the K. A. Wallenberg Foundation
\item[29.] Institut universitaire de France, 75005 Paris, France
\item[30.] Department of Physics and Department of Astronomy, University of Maryland, College Park, MD 20742, USA
\item[31.] Universit\'e Joseph Fourier - Grenoble 1 / CNRS, laboratoire d'Astrophysique de Grenoble (LAOG) UMR 5571, BP 53, 38041 Grenoble Cedex 09, France
\item[32.] Funded by contract ERC-StG-200911 from the European Community
\item[33.] CNRS/IN2P3, Centre d'\'Etudes Nucl\'eaires Bordeaux Gradignan, UMR 5797, Gradignan, 33175, France
\item[34.] Universit\'e de Bordeaux, Centre d'\'Etudes Nucl\'eaires Bordeaux Gradignan, UMR 5797, Gradignan, 33175, France
\item[35.] Dipartimento di Fisica, Universit\`a di Udine and Istituto Nazionale di Fisica Nucleare, Sezione di Trieste, Gruppo Collegato di Udine, I-33100 Udine, Italy
\item[36.] Osservatorio Astronomico di Trieste, Istituto Nazionale di Astrofisica, I-34143 Trieste, Italy
\item[37.] Max-Planck-Institut f\"ur Radioastronomie, Auf dem H\"ugel 69, 53121 Bonn, Germany
\item[38.] Department of Physical Sciences, Hiroshima University, Higashi-Hiroshima, Hiroshima 739-8526, Japan
\item[39.] Agenzia Spaziale Italiana (ASI) Science Data Center, I-00044 Frascati (Roma), Italy
\item[40.] INAF Istituto di Radioastronomia, 40129 Bologna, Italy
\item[41.] Center for Space Plasma and Aeronomic Research (CSPAR), University of Alabama in Huntsville, Huntsville, AL 35899, USA
\item[42.] Instituci\'o Catalana de Recerca i Estudis Avan\c{c}ats (ICREA), Barcelona, Spain
\item[43.] Research Institute for Science and Engineering, Waseda University, 3-4-1, Okubo, Shinjuku, Tokyo, 169-8555 Japan
\item[44.] Istituto Nazionale di Fisica Nucleare, Sezione di Roma ``Tor Vergata", I-00133 Roma, Italy
\item[45.] Department of Physics and Astronomy, University of Denver, Denver, CO 80208, USA
\item[46.] Institute of Space and Astronautical Science, JAXA, 3-1-1 Yoshinodai, Sagamihara, Kanagawa 229-8510, Japan
\item[47.] Hiroshima Astrophysical Science Center, Hiroshima University, Higashi-Hiroshima, Hiroshima 739-8526, Japan
\item[48.] Max-Planck Institut f\"ur extraterrestrische Physik, 85748 Garching, Germany
\item[49.] Institut f\"ur Astro- und Teilchenphysik and Institut f\"ur Theoretische Physik, Leopold-Franzens-Universit\"at Innsbruck, A-6020 Innsbruck, Austria
\item[50.] Space Sciences Division, NASA Ames Research Center, Moffett Field, CA 94035-1000, USA
\item[51.] NYCB Real-Time Computing Inc., Lattingtown, NY 11560-1025, USA
\item[52.] Astro Space Center of the Lebedev Physical Institute, 117810 Moscow, Russia
\item[53.] Astronomical Observatory, Jagiellonian University, 30-244 Krak\'ow, Poland
\item[54.] Department of Chemistry and Physics, Purdue University Calumet, Hammond, IN 46323-2094, USA
\item[55.] Partially supported by the International Doctorate on Astroparticle Physics (IDAPP) program
\item[56.] Consorzio Interuniversitario per la Fisica Spaziale (CIFS), I-10133 Torino, Italy
\item[57.] INTEGRAL Science Data Centre, CH-1290 Versoix, Switzerland
\item[58.] Dipartimento di Fisica, Universit\`a di Roma ``Tor Vergata", I-00133 Roma, Italy
\item[59.] Department of Physics, Royal Institute of Technology (KTH), AlbaNova, SE-106 91 Stockholm, Sweden
\item[60.] School of Pure and Applied Natural Sciences, University of Kalmar, SE-391 82 Kalmar, Sweden
\item[61.] Kwasan and Hida Observatories, Kyoto University, Kyoto 607-8471, Japan
\item[62.] Miyaki-Argenteus Observatory, Miyaki-cho, Saga-ken, Japan
\item[63.] Department of Astronomy, University of California, Berkeley, CA 94720-3411, USA
\item[64.] Department of Astronomy and Astrophysics, Pennsylvania State University, University Park, PA 16802, USA
\item[65.] Hat Creek Observatory, Hat Creek, CA 96040, USA
\item[66.] American Association of Variable Star Observers, Cambridge, MA 02138, USA
\item[67.] Astronomical Institute, Academy of Sciences of the Czech Republic, Ond\v{r}ejov, Czech Republic
\item[68.] Mullard Space Science Laboratory, University College London, Holmbury St. Mary, Dorking, Surrey, RH5 6NT, UK
\item[69.] Sternberg Astronomical Institute, Moscow University, Moscow, Russia
\item[70.] Dipartimento di Fisica "Enrico Fermi" , Universit\`a di Pisa, Pisa I-56127, Italy
\item[71.] Columbia Astrophysics Laboratory, Columbia University, New York, NY 10027, USA
\item[72.] Catholic University of America, Washington, DC 20064, USA

\newpage
\item[*] To whom correspondence should be addressed.\\ 
E-mail: Teddy.Cheung.ctr@nrl.navy.mil (C.C.C.); 
Adam.Hill@obs.ujf-grenoble.fr (A.B.H.);
Pierre.Jean@cesr.fr (P.J.);
srazzaque@ssd5.nrl.navy.mil (S.R.);
Kent.Wood@nrl.navy.mil (K.S.W.)

\end{enumerate}

\newpage
\setcounter{page}{1}

\section*{Supporting Online Material (SOM)}

\section*{Materials and Methods: \Fermi-LAT Data Analysis}

The analysis of the LAT data was performed using the \Fermi\ Science 
Tools v9r15 package available from the \Fermi\ Science Support Center 
(FSSC)\footnote{See the FSSC website for details of the Science Tools: 
http://fermi.gsfc.nasa.gov/ssc/data/analysis/}. The standard onboard 
filtering, event reconstruction, and classification were applied to the 
data \cite{atw09s}, and the high-quality (``Pass 6 diffuse") event class 
is used.  Throughout the analysis, the ``Pass 6 v3 Diffuse'' 
(P6$\_$V3$\_$DIFFUSE) instrument response functions (IRFs) are applied.

Events in the range 0.1--100 GeV were extracted from a 
10\deg$\times$10\deg\ square region of interest (ROI) centered on the 
known location of V407 Cyg. To greatly reduce contamination from the 
Earth albedo photons, we excluded time periods when the 10\deg\ region 
around V407 Cyg was observed at a zenith angle greater than 105\deg\ and 
for observatory rocking angles of greater than 52\deg\ for observations 
after mission elapsed time (MET) 273628805 (rocking angles of greater 
than 43\deg\ are excluded for observations prior to this time).

The {\tt gtlike} likelihood fitting tool was used throughout to perform 
a binned spectral analysis, wherein a spectral-spatial model containing 
point and diffuse sources is created and the parameters obtained from a 
simultaneous maximum likelihood fit to the data.  The model was 
constructed by including the 5 brightest point sources from the 1FGL 
catalog \cite{1fgls} within 15\deg\ of the center of the ROI: 1FGL 
J2021.0+3651, 1FGL J2021.5+4026, 1FGL J2032.2+4127, 1FGL J2030.0+3641, 
and 1FGL J2111.3+4607. The other 1FGL sources within the ROI were not 
considered as they are intrinsically faint and not detectable on the 
timescales of this outburst.  The standard models for the Galactic 
diffuse emission ($gll\_iem\_v{\it 02}.fit$) and isotropic 
backgrounds\footnote{Descriptions of the models are available from the 
FSSC: http://fermi.gsfc.nasa.gov/} currently recommended by the LAT team 
were also incorporated into the model. The first three 1FGL sources 
listed above are known pulsars and were modeled by an exponentially cut 
off power-law, while the remaining two sources were modeled by a single 
power-law with parameters initially set to the values obtained in the 
1FGL catalog \cite{1fgls}. In the fitting, the Galactic diffuse emission 
model was scaled by a single power-law with free normalization and index 
in order to allow for small spectral errors in the model of diffuse 
emission. All source parameters were left free when investigating the 
average spectral behavior over the duration of the entire outburst. The 
parameters of the background point sources were then fixed to their 
average values when performing likelihood analysis on shorter 
timescales.

\subsection*{The average spectral behavior}

A detailed re-analysis of the LAT data on the reported peak flare days 
(13 and 14 March) \cite{che10s} confirms a new $\gamma$-ray source not 
previously reported by the LAT.  Fitting the source with a single 
power-law spectrum yields a flux ($>$100 MeV) = $(1.3 \pm 0.2) \times 
10^{-6}$ photons cm$^{-2}$ s$^{-1}$, and slope, $\Gamma = 2.2 \pm 0.1$, 
with a test statistic (TS, \cite{mat96s}) of 111, where the source 
significance, $\sim\sqrt{TS}$ = 10.5$\sigma$.  Subsequent analysis 
indicated that the source was still detectable on timescales of a day 
through 25 March. Consequently, to investigate the average spectrum and 
maximize the statistical significance of the detection, we define a time 
window extended back to the approximate onset of the optical outburst 
seen in V407 Cyg (10 March 18:00).  To include any late low-level 
emission which would not be detectable on an individual daily basis, the 
window was extended out to 29 March 00:00. This gives a MET range of 
289936803--291513602 and the results of using this defined ``active 
period'' are described below.

Using the {\tt gtfindsrc} tool on this active period gives an improved 
localization over the initially reported one \cite{che10} at (J2000.0) 
R.A. = 315.551\deg, Dec. = 45.737\deg\ ($l = 86.958\deg$, $b= 
-0.513\deg$) with a 95$\%$ error radius of 0.062\deg. This $\gamma$-ray 
position is 0.040\deg\ offset from the optical position of V407 Cyg; 
throughout the remainder of the analysis the nominal optical position of 
V407 Cyg is used.  The previously discussed model was used with the {\tt 
gtlike} tool to identify the average spectral properties of the source 
during this period.  The source was best fit by an exponentially cut off 
power-law model ($dN/dE \propto E^{-\Gamma} e^{-(E/E_{\rm c})}$). The 
flux ($>$100 MeV) obtained is $(4.4 \pm 0.4 \, (\textrm{stat}) \pm 0.2 
\, (\textrm{syst}) ) \times 10^{-7}$ photons cm$^{-2}$ s$^{-1}$, with a 
photon index, $\Gamma = 1.5 \pm 0.2 \, (\textrm{stat}) \pm 0.04 \, 
(\textrm{syst})$, and a cutoff energy, $E_{\rm c} = 2.2 \pm 0.8 \, 
(\textrm{stat}) \pm 0.2 \, (\textrm{syst})$ GeV; see below for 
discussion of how the systematic errors are estimated.  The source is 
detected with a TS = 326.9 ($\sim$18.1$\sigma$).  The cutoff power-law 
was compared to a single power-law model by a likelihood ratio test.  
This gives TS = $-2 \Delta$log(Likelihood) = 23.6 and indicates that the 
addition of the exponential cut off improves the fit at the 4.9$\sigma$ 
level compared to a single power-law model.

A number of effects are expected to contribute to the systematic errors. 
Primarily, these are uncertainties in the effective area and energy 
response of the LAT as well as background contamination. These are 
currently estimated by using outlier IRFs that bracket the nominal ones 
in effective area. These are defined by envelopes above and below the 
P6\_V3\_DIFFUSE IRFs by linearly connecting differences of (10\%, 5\%, 
2\%) at log($E/$MeV) of (2, 2.75, 4), respectively.

\subsection*{Calculating the upper limit}

Using the $\sim$19 months of all-sky monitoring data prior to the onset 
of activity from this source described above it is possible to calculate 
an average flux upper limit for V407 Cyg in this time range. To this end 
a new source model is constructed which comprises all 38 1FGL point 
sources \cite{1fgls} within 15\deg\ of the center of the ROI and the 
standard models for the isotropic and Galactic diffuse emissions. All of 
the 1FGL sources were modeled with single power-law spectra except for 
those known to be pulsars in which case an exponentially cut off 
power-law was applied.  An additional point source is inserted at the 
location of V407 Cyg with an exponentially cut off power-law fixed to 
the average spectral parameters achieved in the fit to its active 
period. The normalization of the source is allowed to be free and a 
maximum likelihood fit performed.  Applying the method of \cite{hel91s} 
the 95$\%$ upper limit on the flux ($>$100 MeV) is calculated to be 
6$\times$10$^{-9}$ photons cm$^{-2}$ s$^{-1}$.

\subsection*{Exploring the evolution of the outburst}

To investigate the evolution of the source flux the data were extracted 
in one-day segments. A likelihood analysis was performed on each of 
these segments using the optimized background source model and an 
exponentially cut off power-law spectrum for the source of interest.  
The spectral model parameters were fixed to the best obtained average 
values and only the source normalization and the diffuse background 
sources were allowed to vary. In each segment the source TS, flux 
($>$100 MeV), and 95$\%$ flux ($>$100 MeV) upper limit were calculated 
(Table~S1). This analysis indicated that the first day in which a 
significant detection for the source was achieved on March 10 and that 
the source was detected (TS$>$9) on daily timescales up to and including 
25 March. To further explore the onset of detectable $\gamma$-ray 
emission around the date of the optical detection of the nova (10 
March), we divided the data into 6-hr bins. This analysis indicated that 
the $\gamma$-ray signal was isolated to the latter half of this day 
(Table~S2).

\subsection*{Search for spectral variability}

We split the \Fermi-LAT observation period into two time segments with 
roughly equivalent statistical significance (TS value) for the detection 
of the source: 10 March 18:00 - 14 March 12:00 (a) and 14 March 12:00 -- 
29 March 00:00 (b). The best fit spectral index and cutoff energy of the 
power-law with exponential cutoff are $\Gamma = 1.6 \pm 0.2$ and $E_{\rm 
c} = (3.0 \pm 0.9)$ GeV, for the first segment (a), and $\Gamma = 1.3 
\pm 0.3$ and $E_{\rm c} = (1.7 \pm 0.6)$ GeV for the second segment (b) 
(Fig.~S1). The flux ($>$100 MeV) varies from $(7.2 \pm 3.1) \times 
10^{-7}$ photons s$^{-1}$ cm$^{-2}$ to (3.3$\pm$0.8) $\times 10^{-7}$ 
photons s$^{-1}$ cm$^{-2}$ between the two periods while the index and 
cutoff energy are statistically equal. For comparison, we fitted the 
total flux of the power law with exponential cutoff with the segment (b) 
data by keeping the index and cutoff energy fixed to the values obtained 
with the best fit of segment (a). It resulted in a difference in 
log(likelihood) of 0.93. This difference corresponds to a significance 
of 1.4$\sigma$ which means that we did not detect any spectral 
variability.


\section*{Parameters of the $\gamma$-ray Emission Models}

We calculate the \p0\ production by $pp$ interactions following the 
prescription of \cite{kam06s}. We use cosmic-ray proton spectra of the 
form, $N_{\rm p} = N_{\rm p,0} \, (W_{\rm p} + m_{\rm p} \, 
c^2)^{-s_{\rm p}} \, e^{-W_{\rm p}/E_{\rm cp}}$ (proton GeV$^{-1}$), 
where $W_{\rm p}$ is the kinetic energy of protons (GeV) and $m_{\rm p}$ 
the proton mass. We fit the normalization ($N_{\rm p,0}$), spectral 
index ($s_{\rm p}$), and the cutoff energy ($E_{\rm cp}$), with the 
\Fermi-LAT data. The \p0\ emissivity was calculated assuming a solar 
metallicity. We apply a corresponding nuclear enhancement factor of 
$\epsilon_{\rm M} =$ 1.84 to the \p0\ emissivity \cite{mor09s}.

The resulting best fit spectral model presented in the main text has a 
spectral index, $s_{\rm p}=2.15^{+0.45}_{-0.28}$ and $E_{\rm cp} = 
32^{+85}_{-8}$ GeV (1$\sigma$ uncertainties). Fig.~S2 shows the 
confidence region of the spectral index and cutoff energy fit to the 
\Fermi-LAT data. Although the uncertainty of the spectral index is 
large, its best fit value corresponds to the canonical slope of 
cosmic-ray protons ($s_{\rm p} \sim 2$) in the first-order Fermi 
acceleration process.  Note that a cutoff energy larger than $\sim$100 
GeV is not excluded for spectral indices larger than $\sim$2.5 (95$\%$ 
confidence level).

The observed $\gamma$-ray flux in the \p0\ model can be reproduced with 
a total number of cosmic-ray protons, $\int N_{\rm p} \, dW_{\rm p}$ = 
$2.0^{+1.1}_{-0.7} \times 10^{45} \, (n_{\rm H}/4 \times 10^{8}~{\rm 
cm}^{-3})^{-1} \, (D/2.7 \,{\rm kpc})^{2}$ in a steady state.  The total 
energy in protons is $\int W_{\rm p} \, N_{\rm p} \, dW_{\rm p}$ = 
$6.9^{+3.6}_{-2.3} \times 10^{42} \, (n_{\rm H}/4 \times 10^{8}~{\rm 
cm}^{-3})^{-1} \, (D/2.7 \,{\rm kpc})^{2}$ ergs.  Here, $n_{\rm H} 
\approx 4 \, n(R)$, is the density of target particles in hydrogen gas 
in the shock region and $D=2.7$~kpc is the adopted distance to V407 Cyg 
\cite{mun90s}.

The inverse Compton and bremsstrahlung emissions in the leptonic model 
are calculated using the method presented in \cite{blu70s}.  In our 
steady state calculation, cosmic-ray electrons in the nova shell 
interact with infrared photons from the red giant (RG; modeled as a 
black-body with a temperature of 2500~K and a radius of 500~$R_{\odot}$) 
at a distance of $\sim$10$^{14}$ cm from the RG.  At this distance, the 
target ion density for the bremsstrahlung process is $n_{\rm H} \approx 
4 \times 10^{8}$ cm$^{-3}$ as in the \p0\ model.  We use cosmic-ray 
electron spectra of the form, $N_{\rm e} = N_{\rm e,0} \, W_{\rm 
e}^{-s_{\rm e}} \, e^{-W_{\rm e}/E_{\rm ce}}$ (electron GeV$^{-1}$), 
where $W_{\rm e}$ is the kinetic energy of electrons (GeV).  The 
normalization factor ($N_{\rm e,0}$), spectral index ($s_{\rm e}$), and 
cutoff energy ($E_{\rm ce}$) are varied to fit the \Fermi-LAT data.

The resulting best fit leptonic spectral model presented in the main 
text (Fig.~3) has a spectral index, $s_{\rm e}=-1.75^{+2.40}_{-0.59}$ 
and $E_{\rm ce} = 3.2^{+2.6}_{-0.1}$ GeV (1$\sigma$ uncertainties) 
(Fig.~S3). Fig.~S4 shows the confidence region of the spectral index and 
cutoff energy fit to the \Fermi-LAT data. It presents two optimal zones 
(confidence level 68$\%$), around $s_{\rm e} = -1.75$ and $s_{\rm e} = 
0.25$.  The best fit electron level leads to a total number of 
cosmic-ray electrons in a steady state of $\int N_{\rm e} \, dW_{\rm e}$ 
= $0.39^{+24.6}_{-0.31} \times 10^{43} \, (D/2.7 \,{\rm kpc})^{2}$ and 
the total energy in electrons of $\int W_{\rm e} \, N_{\rm e} \, dW_{\rm 
e}$ = $0.54^{+34.2}_{-0.43} \times 10^{41} \, (D/2.7 \,{\rm kpc})^{2}$ 
ergs.  Note that the uncertainty in the electron spectral index is 
rather large, and the canonical slope ($s_{\rm e} \sim 2$) expected to 
arise from the first-order Fermi acceleration process cannot be excluded 
(99$\%$ confidence level).

The difference in the log(likelihood) value for the best fit leptonic 
model and that of the best fit \p0\ model is $\sim$0.05.  This 
difference corresponds to a significance of $\sim$0.3$\sigma$ and 
neither model is statistically preferred over the other.

\section*{Optical and Infrared Photometry of V407 Cyg}

The optical nova of V407 Cyg was discovered by Nishiyama and Kabashima 
\cite{nis10s} using an unfiltered CCD image obtained on 10.797 March 
2010 UT at the Miyaki Argenteus Observatory in Japan with a 105-mm f/4.0 
lens telescope. The pre-outburst image from 7.859 March shows the source 
2.5 magnitudes fainter. There are uncertainties in the actual epoch of 
the nova due to the three-day gap in the observations. Previous images 
obtained with the same equipment dating back to April 2008 (Fig.~S5) 
show slow light variations with the total amplitude $\sim$3~mag, which 
may be attributed to a combination of the Mira-type pulsations of the 
red giant and activity of the white dwarf.

The bright outburst triggered follow-up photometric observations with a
range of instruments and the observations shown in the main paper are
summarized here (see Table~S3). Many CCD measurements were contributed by
enthusiasts from the American Association of the Variable Star Observers
(AAVSO), and those who contributed $V$ and $R_{\rm C}$ band CCD photometry
presented here were:
G.~Belcheva (Bulgaria),
S.~Dvorak,
M.~Halderman,
G.~Sjoberg,
D.~Trowbridge (USA),
T.~Kantola,
M.~Luostarinen,
A.~Oksanen,
J.~Virtanen (Finland),
D.~Lane (Canada),
S.~O'Connor (Bermuda),
S.~Padovan (Spain), and
A.~Smirnov (Russia).
Observations from the Kwasan Observatory and KANATA telescope are also
reported. Small systematic discrepancies between CCD measurements   
obtained with different telescopes (resulting from different choices of
comparison stars and mismatches between instrumental and standard
bandpass) were compensated when necessary by using the well sampled
AAVSO Bright Star Monitor (BSM) light curve as the primary data set. The
BSM photometry used an ensemble of comparison stars in the 8--9 mag
range calibrated using the Landolt standards \cite{lan92s}.

Fig.~S5 shows the combined optical light curve in $V$ and $R_{\rm C}$ 
bands and the unfiltered CCD measurements by K.~Nishiyama and 
F.~Kabashima. A subset of these data are shown in the main paper. The 
magnitude zero point for the unfiltered measurements was calculated 
assuming $R_{\rm C}=6.571$ for the comparison star HIP 103871 as 
indicated by the BSM photometry, resulting in the peak magnitude 
$\sim$0.5 mag brighter than the value reported in the discovery 
\cite{nis10s}. The $V$-band light curve peak of 7.8 mag on 11 March 2010 
20:28 communicates a larger brightening, in comparison with the 
(sparsely covered) historical magnitudes of 14--15, than is apparent in 
the unfiltered photometric data.

V407 Cyg was also observed in the infrared nine and twenty-five days 
after the nova explosion using the InSb-photometer \cite{nad86s} at the 
1.25-m telescope of the Crimean Laboratory (Sternberg Astronomical 
Institute). The $JHKLM$ band magnitudes observed are typical for this 
system around maximum of the Mira pulsation cycle (although with 
slightly bluer colors), indicating that the RG was still dominating in 
the infrared \cite{kol03s}. Assuming the Galactic extinction ($E_{\rm 
B-V}=0.57$) adopted in this paper \cite{mun90s}, we converted magnitudes 
to flux densities \cite{bes88s} and fitted the infrared spectra with 
Planck's law, which gave a RG temperature of 2500~K at both epochs.

\section*{Optical Spectroscopic Measurements}

The ejecta velocity, $v_{\rm ej} = 3200 \pm 345$ km s$^{-1}$, quoted in 
the main paper was taken as the half width zero intensity of the 
H$\alpha$ line measured in a spectrum obtained at the Castanet-Tolosan 
Observatory and made available online by 
C.~Buil\footnote{http://astrosurf.com/aras/V407Cyg/v407cyg.htm}.  The 
measurement was done by fitting a 6th order polynomial to the continuum 
only in the range $\lambda$6420--6655\Angst, and subtracting from the 
observed spectrum.

Optical spectra of V407 Cyg obtained with the Nordic Optical Telescope 
(program P40-423) with resolution 2.2~km s$^{-1}$ (resolving power = 
67000) showed that the O I] $\lambda$6363\Angst\ line displayed a narrow 
emission peak at $-$54~km s$^{-1}$ (Fig.~S6) on 31 March with FWHM = 
26~km s$^{-1}$, as well as strong absorption on Na I 
$\lambda\lambda$5889, 5895\Angst\ (components at $-$58 and $-$51~km 
s$^{-1}$). Ca I $\lambda$4226.7\Angst\ shows a single emission 
component, centered at $-$55~kms$^{-1}$, with FWHM of 7~km s$^{-1}$ that 
is clearly from the stellar wind.  The Balmer lines show strong, broad P 
Cyg profiles centered at $-$60~km s$^{-1}$, indicating absorption 
through the wind of the RG. The permitted and forbidden Fe-peak lines 
(i.e., Fe II) display narrow components and the strongest permitted 
lines display P Cyg profiles with absorption component displacements 
indicating a RG wind velocity, $v_{\rm w}\approx 10$~km s$^{-1}$.


\newpage

\begin{figure}[htbp]
  \begin{center}
     \includegraphics[width=16cm]{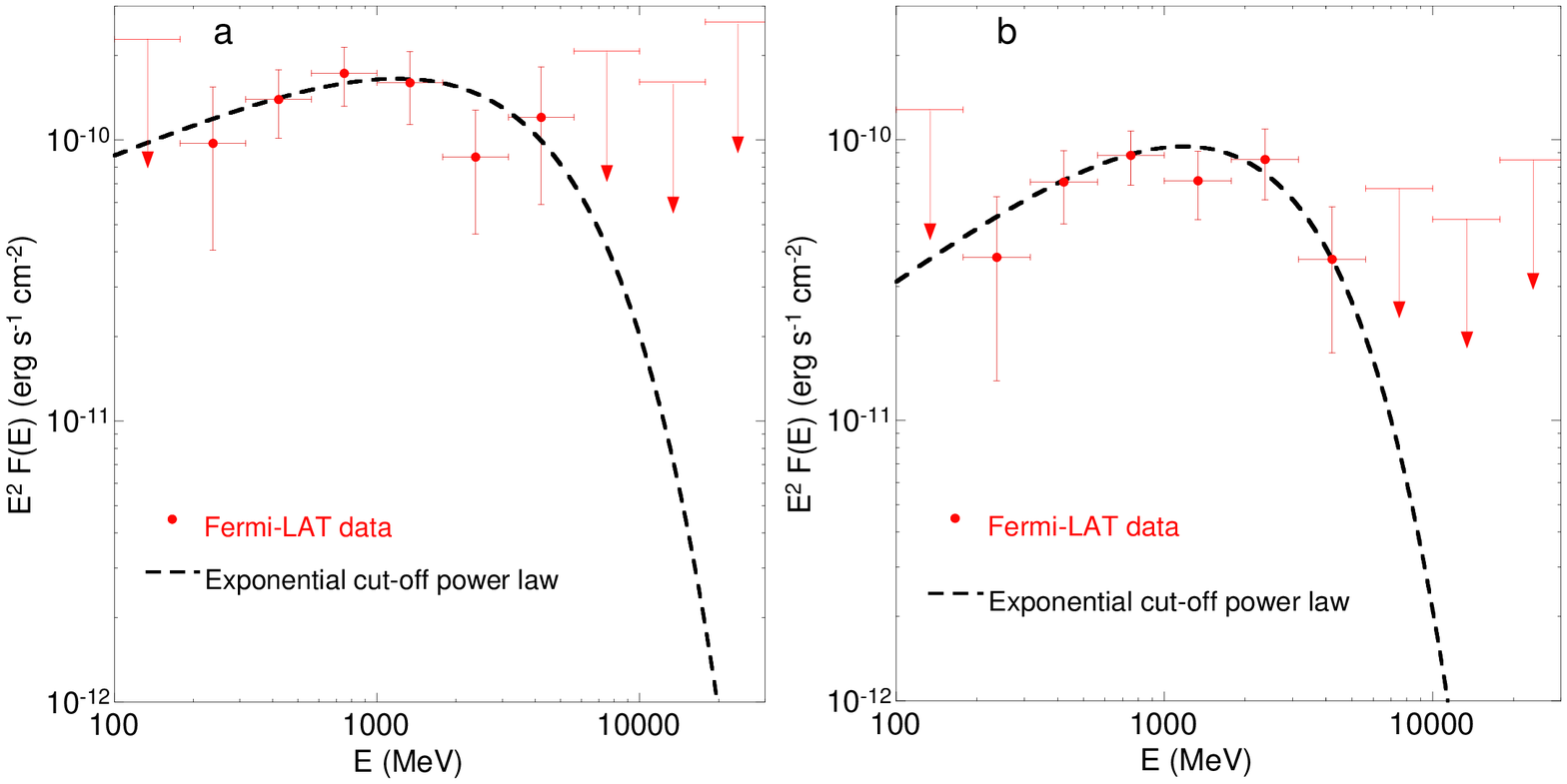}
  \end{center} {\bf Fig.~S1.} Spectral energy distribution of V407 Cyg 
in MeV/GeV $\gamma$-rays measured by the \Fermi-LAT over the period from 
10 March 18:00 to 14 March 12:00 (a) and 14 March 12:00 to 29 March 
00:00 2010 (b). Horizontal bars indicate energy ranges, vertical bars 
indicate 1$\sigma$ statistical errors, and arrows indicate 2$\sigma$ 
upper limits. The best fit of a phenomenological power-law with 
exponential cutoff (dashed line) is presented for the two periods. 
\end{figure}

\begin{figure}[htbp]
  \begin{center}
    \includegraphics[width=12cm,angle=90]{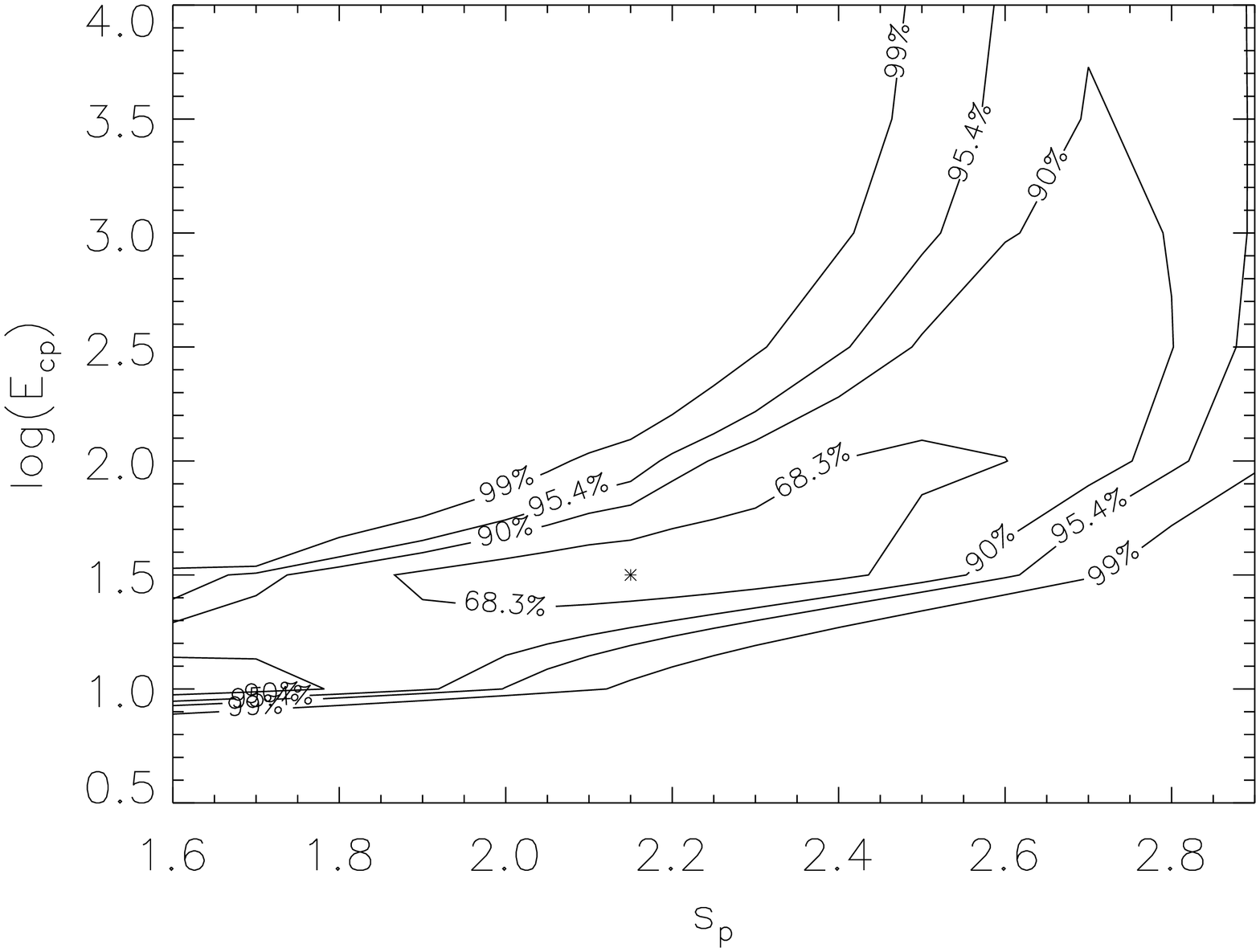}
  \end{center} {\bf Fig.~S2.} Confidence region of the spectral slope 
($s_{\rm p}$) and logarithm of the cutoff energy ($E_{\rm cp}$ in GeV) 
fit of the \Fermi-LAT data for V407 Cyg with the cosmic-ray proton 
spectrum. The star indicates the best-fit values. \end{figure}

\begin{figure}[htbp]
  \begin{center}
    \includegraphics[width=15cm]{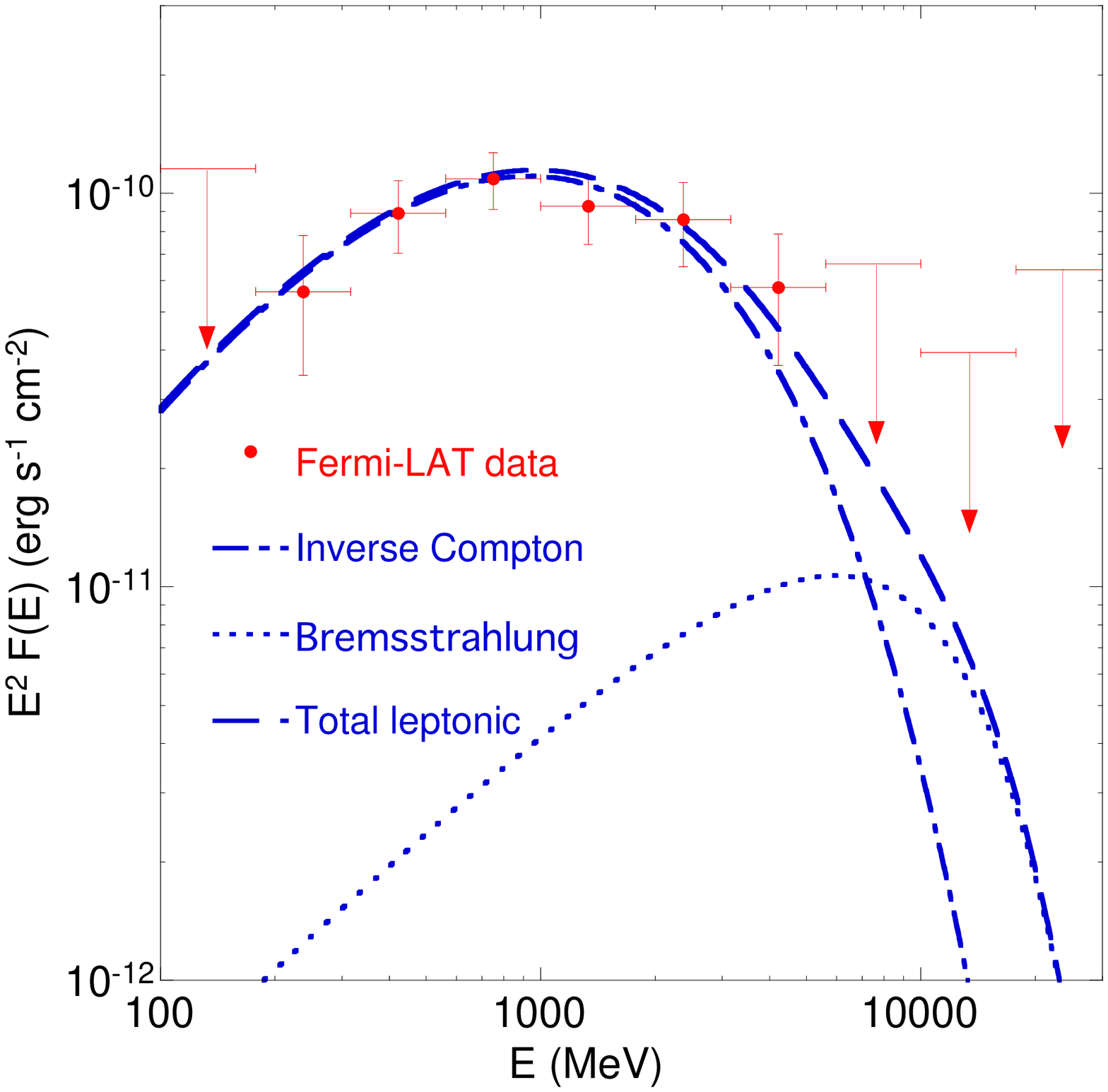}
  \end{center} {\bf Fig.~S3.} SED of V407 Cyg in MeV/GeV $\gamma$-rays 
measured by the \Fermi-LAT over the period 10 March 18:00 -- 29 March 
00:00 2010. Vertical bars indicate 1$\sigma$ statistical errors, arrows 
indicate 2$\sigma$ upper limits, and horizontal bars indicate energy 
ranges. The best-fit leptonic model with the separate contributions from 
the IC (dot-dashed line) and bremsstrahlung (dotted line) spectra 
indicated, as well as their total (dashed line). \end{figure}

\begin{figure}[htbp]
  \begin{center}
    \includegraphics[width=12cm,angle=90]{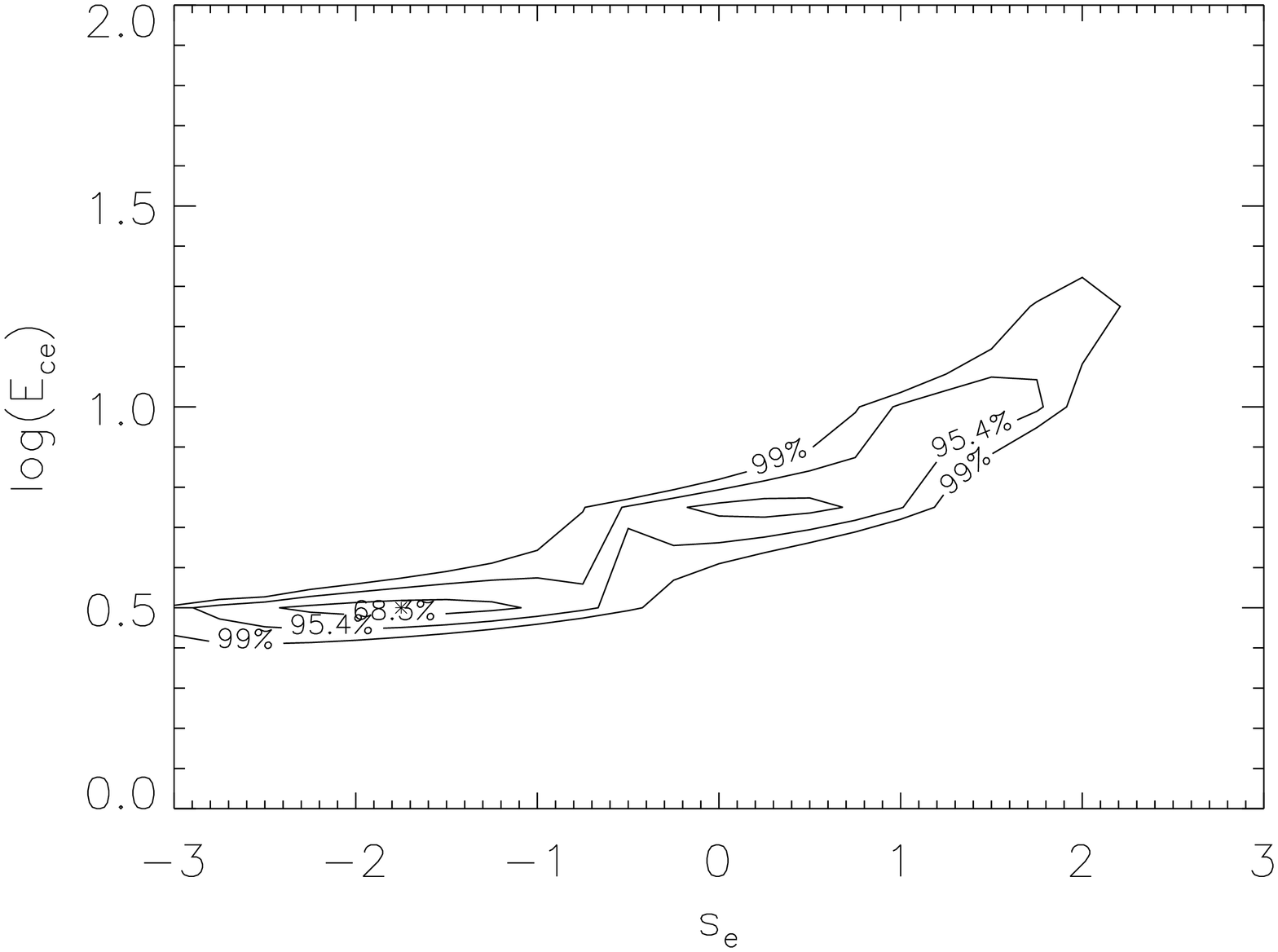}
  \end{center} {\bf Fig.~S4.} Confidence region of the spectral slope 
($s_{\rm e}$) and logarithm of the cutoff energy ($E_{\rm ce}$ in GeV) 
fit of the \Fermi-LAT data for V407 Cyg with the cosmic-ray electron 
spectrum. The star indicates the best-fit values. \end{figure}

\begin{figure}[htbp]   
  \begin{center}
    \includegraphics[width=16cm]{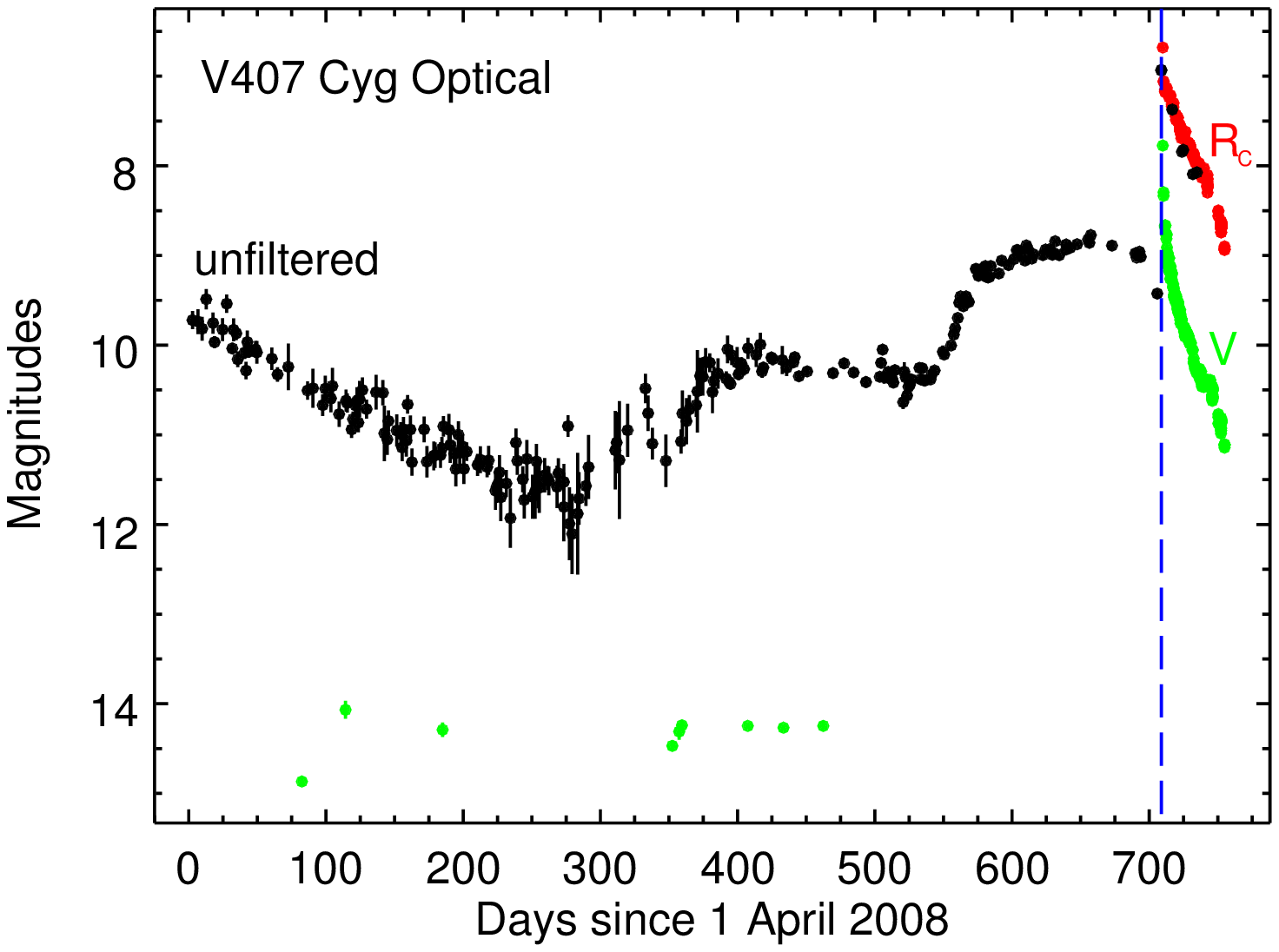}
  \end{center} {\bf Fig.~S5.} Optical light curve of V407 Cyg extending 
back to April 2008. The blue vertical line indicates the epoch of the 
nova discovery on 10 March 2010. \end{figure}

\begin{figure}[htbp] 
  \begin{center}
    \includegraphics[width=16cm]{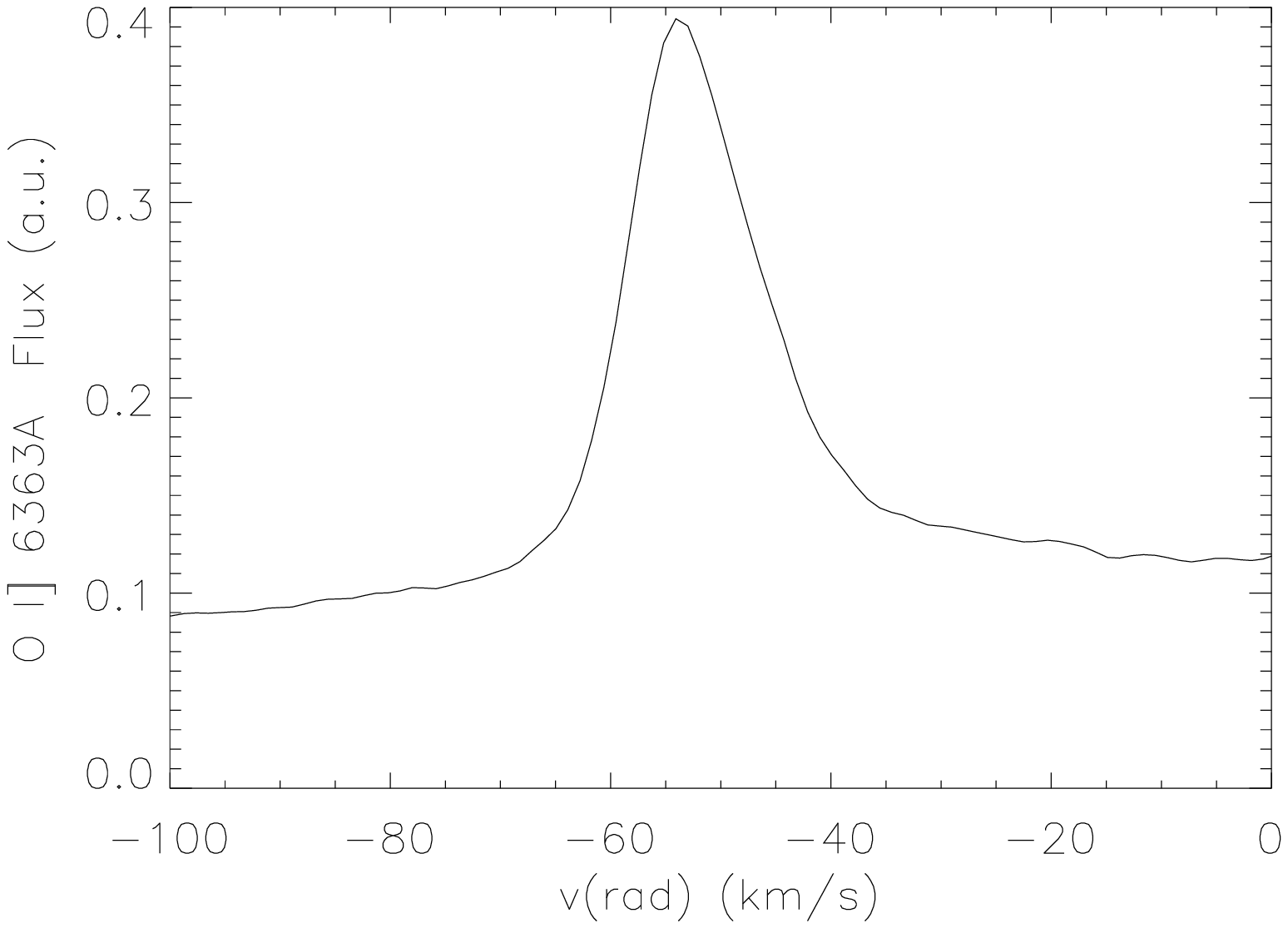}
  \end{center} {\bf Fig.~S6.} The O I] $\lambda$6303\Angst\ line in the 
heliocentric radial velocity observed in V407 Cyg with the Nordic 
Optical Telescope on 31 March 2010. \end{figure}


\begin{table}
\footnotesize
\begin{center}
\begin{tabular}{lcc}
\hline\hline
\multicolumn{1}{l}{Days since} &
\multicolumn{1}{c}{TS} &
\multicolumn{1}{c}{Flux ($>$100 MeV)} \\
\multicolumn{1}{l}{10 March 2010} &
\multicolumn{1}{c}{} &
\multicolumn{1}{c}{[$10^{-7}$ photons cm$^{-2}$ s$^{-1}$]} \\
\hline
--4.5 & 1.5 & $<$4.4 \\
--3.5 & 0.0 & $<$2.5 \\
--2.5 & 0.0 & $<$2.2 \\
--1.5 & 2.2 & $<$3.6 \\
--0.5 & 0.0 & $<$1.4 \\
0.5 & 18.9 & 4.3 $\pm$ 1.6 \\
1.5 & 56.9 & 7.2 $\pm$ 1.7 \\
2.5 & 12.0 & 3.7 $\pm$ 1.5 \\
3.5 & 68.4 & 9.0 $\pm$ 1.9 \\
4.5 & 47.8 & 8.7 $\pm$ 2.0 \\
5.5 & 27.8 & 7.3 $\pm$ 2.0 \\
6.5 & 16.7 & 3.3 $\pm$ 1.3 \\
7.5 & 9.7 & 3.7 $\pm$ 1.8 \\
8.5 & 11.7 & 2.7 $\pm$ 1.2 \\
9.5 & 42.2 & 7.6 $\pm$ 2.0 \\
10.5 & 11.6 & 2.6 $\pm$ 1.2 \\
11.5 & 15.9 & 3.7 $\pm$ 1.6 \\
12.5 & 0.5 & $<$2.8 \\
13.5 & 7.4 & $<$6.9 \\
14.5 & 19.7 & 4.9 $\pm$ 1.7 \\
15.5 & 13.3 & 4.9 $\pm$ 1.8 \\
16.5 & 0.0 & $<$3.5 \\
17.5 & 4.8 & $<$5.7 \\
18.5 & 0.0 & $<$1.8 \\
19.5 & 2.4 & $<$3.5 \\
20.5 & 0.2 & $<$3.2 \\
21.5 & 0.9 & $<$3.7 \\
22.5 & 0.2 & $<$4.2 \\
23.5 & 0.0 & $<$3.1 \\
24.5 & 0.0 & $<$2.0 \\
25.5 & 2.1 & $<$4.3 \\
26.5 & 0.0 & $<$3.3 \\
27.5 & 1.9 & $<$3.2 \\
28.5 & 0.0 & $<$1.8 \\
29.5 & 0.0 & $<$2.8 \\
30.5 & 0.2 & $<$3.2 \\
31.5 & 0.0 & $<$1.7 \\
32.5 & 0.3 & $<$2.2 \\
33.5 & 2.1 & $<$4.7 \\
34.5 & 0.0 & $<$2.2 \\
35.5 & 3.1 & $<$5.5 \\
36.5 & 0.0 & $<$1.9 \\
37.5 & 1.6 & $<$3.5 \\
38.5 & 0.2 & $<$2.9 \\
\hline
\end{tabular}
\normalsize
\label{table-gamma}
\end{center}
{\bf Table~S1.} Daily LAT $\gamma$-ray test statistic (TS) and flux 
values (95$\%$ confidence upper limits when TS$<$9) presented in the 
main paper. The dates indicated are the centers of the one-day bins. 
\end{table}

\begin{table}
\begin{center}
\begin{tabular}{ccc}
\hline\hline
\multicolumn{1}{c}{Time} &
\multicolumn{1}{c}{TS} &
\multicolumn{1}{c}{Flux ($>$100 MeV)} \\
\multicolumn{1}{c}{Interval} &
\multicolumn{1}{c}{} &
\multicolumn{1}{c}{[$10^{-7}$ photons cm$^{-2}$ s$^{-1}$]} \\
\hline
00h-06h & 0.0 & $<$3.9 \\
06h-12h & 0.0 & $<$6.4 \\
12h-18h & 8.0 & 6.2 $\pm$ 3.7\\
18h-24h & 32.0 & 17.4 $\pm$ 5.3 \\ 
\hline
\end{tabular}
\label{table-gamma6}
\end{center}
{\bf Table~S2.} Six-hour $\gamma$-ray test statistic (TS) and flux 
values (or 95$\%$ confidence upper limits) for 10 March. Note that a 
flux is reported for the 12h-18h bin despite showing a TS$<$9 as adopted 
throughout; the corresponding upper limit was $<$15.1$\times$$10^{-7}$ 
photons cm$^{-2}$ s$^{-1}$. \end{table}

\begin{table}
\begin{center}
\begin{tabular}{lcc}
\hline\hline
\multicolumn{1}{c}{Telescope} &
\multicolumn{1}{c}{Filter(s)} &
\multicolumn{1}{c}{Camera} \\
\hline
Astrokolkhoz Observatory AAVSO BSM 60-mm, USA  & $VR_{\rm C}$ & SBIG ST-8XME \\
Kwasan Observatory 250-mm, Japan & $VR_{\rm C}$ & SBIG ST-7XME \\
Hiroshima University KANATA 1.5-m, Japan & $V$ & TRISPEC \cite{wat05s} \\
Miyaki Argenteus Observatory 105-mm, Japan & unfiltered & SBIG STL6303E \\
\hline
\end{tabular}
\label{table-optical}
\end{center}
{\bf Table~S3.} Instruments used for the optical photometric 
monitoring observations (Fig.~S5).
\end{table}


\begin{thebibliography}{99}
\bibitem {onset} The window was extended by three days to include any
late low-level emission which would not be detectable on any
individual daily bin.
\end{thebibliography}
\end{document}